\documentclass[prd,superscriptaddress,amsfonts,amssymb,amsmath,showpacs,twocolumn]{revtex4-2}
\usepackage{bm}
\usepackage{amsfonts}
\usepackage{latexsym}
\usepackage{graphicx}
\usepackage{amsmath}
\usepackage{palatino}
\usepackage{mathpazo}
\usepackage{textcomp}
\usepackage{float}
\usepackage{booktabs}
\usepackage{dcolumn}
\usepackage{booktabs}
\usepackage{multirow}
\usepackage{hyperref}
\hypersetup{colorlinks,citecolor=blue}
\usepackage{amsmath}
\usepackage{xcolor}
\usepackage{orcidlink}
\usepackage{epsfig}
\usepackage{caption}
\usepackage{subcaption}
\usepackage{commath}
\def\jnl@style{\it}
\def\aaref@jnl#1{{\jnl@style#1}}

\def\aaref@jnl#1{{\jnl@style#1}}

\def\aj{\aaref@jnl{AJ}}                   % Astronomical Journal
\def\apj{\aaref@jnl{ApJ}}                 % Astrophysical Journal
\def\apjl{\aaref@jnl{ApJ}}                % Astrophysical Journal, Letters
\def\apjs{\aaref@jnl{ApJS}}               % Astrophysical Journal, Supplement
\def\apss{\aaref@jnl{Ap\&SS}}             % Astrophysics and Space Science
\def\aap{\aaref@jnl{A\&A}}                % Astronomy and Astrophysics
\def\aapr{\aaref@jnl{A\&A~Rev.}}          % Astronomy and Astrophysics Reviews
\def\aaps{\aaref@jnl{A\&AS}}              % Astronomy and Astrophysics, Supplement
\def\mnras{\aaref@jnl{Mon.~Not.~Roy.~Astron.~Soc.}}             % Monthly Notices of the RAS
\def\prd{\aaref@jnl{Phys.~Rev.~D}}        % Physical Review D
\def\prc{\aaref@jnl{Phys.~Rev.~C}}  % Physical Review C
\def\prl{\aaref@jnl{Phys.~Rev.~Lett.}}    % Physical Review Letters
\def\qjras{\aaref@jnl{QJRAS}}             % Quarterly Journal of the RAS
\def\skytel{\aaref@jnl{S\&T}}             % Sky and Telescope
\def\ssr{\aaref@jnl{Space~Sci.~Rev.}}     % Space Science Reviews
\def\zap{\aaref@jnl{ZAp}}                 % Zeitschrift fuer Astrophysik
\def\nat{\aaref@jnl{Nature}}              % Nature
\def\aplett{\aaref@jnl{Astrophys.~Lett.}} % Astrophysics Letters
\def\apspr{\aaref@jnl{Astrophys.~Space~Phys.~Res.}} % Astrophysics Space Physics Research
\def\physrep{\aaref@jnl{Phys.~Rep.}}      % Physics Reports
\def\physscr{\aaref@jnl{Phys.~Scr}}       % Physica Scripta
\def\commat{\aaref@jnl{Comm.~Math.~Phys.}}              % Communications in Mathematical Physics
\def\science{\aaref@jnl{Science}}               % Science
\def\cqg{\aaref@jnl{Classical Quant.~Grav.}}            % Classical and Quantum Gravity
\def\jpcs{\aaref@jnl{JPCS}}                                     % Journal of Physics Conference Series
\def\ijmpd{\aaref@jnl{Int.~J.~Mod.~Phys.~D}}                    % International Journal of Modern Physics D
\def\grg{\aaref@jnl{Gen.~Relat.~Gravit.}}               % General Relativity and Gravitation
\def\rpp{\aaref@jnl{Rep.~Prog.~Phys.}}          % Reports on Progress in Physics
\def\npa{\aaref@jnl{Nucl.~Phys.~A}}        % Nuclear Physics A
\def\lrr{\aaref@jnl{Living Rev.~Rel.}}                   % Living reviews in relativity
\def\jcap{\aaref@jnl{J.~Cosmology Astropart.~Phys.}}    % Journal of cosmology and astroparticle physics
\def\rmp{\aaref@jnl{Rev.~Mod.~Phys.}}   %Reviews of modern physics
\def\epjc{\aaref@jnl{Eur.~Phys.~J.~C}}

\allowdisplaybreaks[1]
\begin{document}

\color{black}       %% For one column

\title{Observational constraints on a generalized equation of state model}

\author{M. Koussour\orcidlink{0000-0002-4188-0572}}
\email[Email: ]{pr.mouhssine@gmail.com}
\affiliation{Department of Physics, University of Hassan II Casablanca, Morocco.}

\author{S. Bekov\orcidlink{0000-0002-8846-0604}}
\email[Email: ]{ss.bekov@gmail.com \textcolor{black}{(corresponding author)}}
\affiliation{Department of General and Theoretical Physics, L.N. Gumilyov Eurasian National University, Astana 010008, Kazakhstan.}
\affiliation{Kozybayev University,  Petropavlovsk, 150000, Kazakhstan.}

\author{A. Syzdykova}
\email[Email: ]{Syzdykova_am@mail.ru}
\affiliation{Department of General and Theoretical Physics, L.N. Gumilyov Eurasian National University, Astana 010008, Kazakhstan.}

\author{S. Muminov\orcidlink{0000-0003-2471-4836}}
\email[Email: ]{sokhibjan.muminov@gmail.com}
\affiliation{Mamun University, Bolkhovuz Street 2, Khiva 220900, Uzbekistan.}

\author{I. Ibragimov}
\email[Email: ]{i.ibragimov@kiut.uz}
\affiliation{Kimyo International University in Tashkent, Shota Rustaveli street 156, Tashkent 100121, Uzbekistan.}

\author{J. Rayimbaev\orcidlink{0000-0001-9293-1838}}
\email[Email: ]{javlon@astrin.uz}
\affiliation{Institute of Fundamental and Applied Research, National Research University TIIAME, Kori Niyoziy 39, Tashkent 100000, Uzbekistan.}
\affiliation{University of Tashkent for Applied Sciences, Str. Gavhar 1, Tashkent 100149, Uzbekistan.}
\affiliation{Urgench State University, Kh. Alimjan Str. 14, Urgench 221100, Uzbekistan}
\affiliation{Shahrisabz State Pedagogical Institute, Shahrisabz Str. 10, Shahrisabz 181301, Uzbekistan.}

%%%%%%%%%%%%%%%%%%%%%%%%%%%%%%%%%%%%  DATE  %%%%%%%%%%%%%%%%%%%%%%%%%%%%%%%%%%%%
\date{\today}

\begin{abstract}
We investigate the cosmological implications of a generalized total equation of state (EoS) model by constraining its parameters using observational datasets to effectively characterize the universe's expansion history and its dynamic properties. We introduce three parameters: $\alpha$, $\beta$, and $n$ to capture the EoS behavior across different evolutionary phases. Our analysis indicates that at high redshifts ($z \gg 1$), the EoS approaches a matter- or radiation-dominated regime, transitioning to a dark energy-dominated phase as $z \to -1$, where it tends towards a constant value $\alpha$. Using a Markov Chain Monte Carlo (MCMC) method, we analyze a combined dataset that includes 31 data points from $H(z)$ and 1701 data points from the Pantheon+ dataset. The results reveal a smooth transition from deceleration to acceleration in the universe's expansion, with current EoS values suggesting quintessence-like behavior. The model aligns with observations and indicates that dark energy is dynamically evolving rather than acting as a cosmological constant. Furthermore, energy conditions and stability analyses highlight the nature and future of dark energy. This parametrized EoS model thus offers a robust framework for understanding the complexities of dark energy and the evolution of the cosmos.

\textbf{Keywords:} dark energy, cosmological models, equation of state, MCMC, quintessence, and energy conditions.
\end{abstract}

\maketitle

\section{Introduction}
\label{sec1}

Observational evidence indicates that the Universe experienced two separate epochs of cosmic acceleration. The first, known as inflation, commenced shortly after the Big Bang and played a critical role in shaping the large-scale structure of the cosmos. This rapid expansion phase helped resolve several fundamental issues in cosmology, such as the horizon and flatness problems, by exponentially stretching the fabric of spacetime \cite{Linde}. Following this initial period, the Universe entered a protracted phase of cosmic acceleration, which began approximately 6 billion years after the Big Bang and continues to this day. This latter phase, often attributed to the influence of dark energy (DE), has profound implications for our understanding of the Universe's fate. Observational evidence, including measurements from type Ia supernovae (SNe)  \cite{Riess, Perlmutter}, the cosmic microwave background (CMB) \cite{R.R., Z.Y.}, the baryon acoustic oscillations (BAO) \cite{D.J., W.J.}, large-scale structure (LSS) surveys \cite{T.Koivisto, S.F.}, and the latest results from the Planck collaboration \cite{Planck2020}, points toward a dynamic form of DE that varies over time. This challenges the traditional cosmological constant ($\Lambda$) model, which represents approximately 70\% of the Universe’s current energy budget and is characterized by a fixed negative pressure. Further constraints on the nature of DE have been confirmed by measurements of large-scale clustering \cite{LSC1,LSC2}, cosmic age \cite{CA}, weak lensing \cite{WL}, and gamma-ray bursts \cite{GRB1,GRB2}. The equation of state (EoS) parameter for DE, $\omega$, which defines the ratio of pressure $p$ to energy density $\rho$, must stay below $-1/3$ in order for late-time cosmic acceleration to occur.

Several DE models have been proposed based on current observational data. The most basic model is called $\Lambda$CDM ($\Lambda$ Cold Dark Matter). The cosmological constant $\Lambda$ in this model is used to describe DE. It is defined by a constant energy density across time and $\omega_{\Lambda} = -1$, which means that the energy density $\rho_{\Lambda}$ is equal to the negative of the pressure $p_{\Lambda}$. As well as explaining many important aspects of the large-scale structure and late-time acceleration of the cosmos, $\Lambda$CDM has two serious theoretical problems. First, the fine-tuning problem arises because the observed value of the cosmological constant is many orders of magnitude smaller than what is predicted by quantum field theories, raising questions about why it is so finely tuned to a specific value. Second, the cosmological coincidence problem refers to the seemingly coincidental timing of the present epoch, during which the energy densities of DE and matter are comparable, even though they evolve differently over time. This raises the question of why we happen to observe the universe during this specific phase when both components are of similar magnitude, despite their different evolutionary trajectories \cite{dalal/2001, weinberg/1989}. These issues have prompted the exploration of alternative models of DE that allow for a dynamic equation of state, such as quintessence, k-essence, or phantom energy \cite{Qu3,Qu4,Phantom1,Phantom2,T.Chiba,C.Arm.,Kamenshchik,M.C.,A.Y.}, to better account for the universe's expansion history and potentially address the shortcomings of the $\Lambda$CDM model. In recent years, there has been a growing interest in modified gravity theories as an alternative explanation for the universe's accelerated expansion. Instead of introducing a separate DE component, these theories propose modifications to the Einstein-Hilbert action by incorporating functions of various curvature invariants, such as the Riemann tensor, Ricci tensor, or Weyl tensor. Notable examples include $f(R)$ gravity, where $R$ represents the curvature scalar \cite{fR1,fR2}; $f(T)$ gravity, based on the torsion scalar $T$ \cite{fT1,fT2}; and $f(Q)$ gravity, which relies on the non-metricity scalar $Q$ \cite{fQ1,fQ2,fQ3,fQ4,fQ5,fQ6,fQ7}.

To understand more about the features of DE and its effects on cosmic evolution, researchers have developed various theoretical frameworks, one of which is the parametrized EoS model \cite{Mukherjee}. This model allows for the exploration of the dynamic properties of DE through a set of parameters that describe its behavior across different evolutionary phases. By constraining these parameters using observational datasets, we can investigate potential deviations from standard cosmological models, such as the $\Lambda$CDM paradigm, and enhance our understanding of the underlying physics driving cosmic acceleration. Numerous parametrizations of the EoS for DE have been proposed in the literature to capture its dynamic behavior and evolution over time. One widely used form is the linear parametrization, given by $\omega(z)=\omega_{0}+\omega_{1} z$, where $\omega_0$ represents the present EoS value, and $\omega_1$ accounts for its variation with redshift \cite{Lin}. The Chevallier-Polarski-Linder (CPL) parametrization, expressed as $\omega(z)=\omega_{0}+\omega_{1} \frac{z}{1 + z}$, provides another popular framework for exploring EoS evolution \cite{CPL1,CPL2}. Other models include exponential forms, such as $\omega(a)=\omega_{0} \exp{(a-1)}$ (where $a=\frac{1}{1+z}$ is the scale factor), which allow for controlled rates of evolution \cite{Exp}, and polynomial parametrizations, which can fit data flexibly by incorporating higher-order terms \cite{Poly}. The Jassal-Bagla-Padmanabhan (JBP) model $\omega(z)=\omega_{0}+\omega_{1}\frac{z}{(1+z)^2}$ captures the transition from radiation or matter domination to DE domination \cite{JBP}, while generalized parametrizations, such as $\omega(z)=\omega_{0}+\omega_{1}\frac{z}{(1+z)^n}$ \cite{JBPn}, include additional parameters to modify EoS evolution. Logarithmic forms, like $\omega(z)=\omega_{0}+\omega_{1}(\frac{\ln(2+z)}{1+z}-\ln2)$ \cite{MZ}, are also utilized to represent specific dynamics. 

In the motivation for our current model, we build upon existing parameterizations of the EoS for DE that effectively describe the universe's accelerated expansion. A. Mukherjee \cite{Mukherjee} explored a parameterization of the effective or total EoS where DE behaves similarly to the cosmological constant in the far future. This parameterization, given by $ \omega(z) = \frac{-1}{1 + \beta (1 + z)^n}$ allows for a dynamic evolution of the EoS while still converging to $\omega = -1$ at late times, consistent with the standard $\Lambda$CDM model. Our model introduces the parameter $\alpha$, which generalizes this approach by providing more flexibility in describing deviations from the cosmological constant and capturing potential dynamical behavior in the DE component across different cosmic epochs. This additional freedom enables us to explore more complex scenarios, including transitions in the DE behavior and possible deviations from $\Lambda$CDM, as suggested by observational data. In this work, we will analyze the cosmological implications of a parametrized total EoS model by leveraging observational data, thereby elucidating the transition from deceleration to acceleration and its significance for the ultimate fate of the Universe.

The paper is organized as follows: Sec. \ref{sec2} presents the theoretical framework of the generalized EoS model. In Sec. \ref{sec3}, we detail the observational datasets used, including the $H(z)$ and Pantheon+ datasets, along with the methodology for determining the best-fit values of the model parameters through Markov Chain Monte Carlo (MCMC) analysis. The results of the statistical analysis and the derived cosmological parameters are discussed in Sec. \ref{sec4}. In Sec. \ref{sec5}, we examine the energy conditions and stability to validate our generalized EoS model. Finally, we summarize our findings and discuss their implications in Sec. \ref{sec6}.

\section{Theoretical Framework}
\label{sec2}

\subsection{Background cosmology}

The Friedmann-Lemaître-Robertson-Walker (FLRW) metric is a fundamental solution in cosmology that describes a homogeneous, isotropic universe. It is derived from Einstein's field equations in GR under the assumption that the universe exhibits the same properties at every point (homogeneity) and looks the same in every direction (isotropy). These assumptions are supported by observations, such as the CMB radiation \cite{R.R., Z.Y.}, and form the basis for modern cosmological models. The line element for the FLRW metric is given by \cite{Ryden}
\begin{equation}
    ds^2 = -c^2 dt^2 + a^2(t) \left[ \frac{dr^2}{1 - k r^2} + r^2 \left( d\theta^2 + \sin^2 \theta \, d\phi^2 \right) \right],
\end{equation}
where $t$ is the cosmic time, $r$, $\theta$, and $\phi$ are the comoving spherical coordinates. The FLRW metric assumes that the geometry of space at any given time is that of a 3-dimensional maximally symmetric space, which can be either spherical, flat, or hyperbolic, depending on the sign of the spatial curvature of the universe, which can take values $k = 0$ (flat universe), $k = 1$ (closed universe with positive curvature), or $k = -1$ (open universe with negative curvature). In many cosmological studies, the universe is often assumed to be flat ($k = 0$) based on observational evidence from the CMB and large-scale structure surveys and the predictions of inflationary models \cite{R.R., Z.Y.,D.J., W.J.,T.Koivisto, S.F.}. In addition, the scale factor $a(t)$ describes the expansion of the universe as a function of time. It is normalized such that $a(t_0) = 1$ at the present time $t_0$. The rate of change of the scale factor is characterized by the Hubble parameter $H(t)$, defined as $H(t) = \frac{\dot{a}(t)}{a(t)}$, where $\dot{a}(t)$ represents the time derivative of the scale factor. The Hubble parameter plays a crucial role in cosmology as it governs the rate at which the universe expands. In the present epoch, the value of the Hubble parameter is denoted as $H_0$ and is referred to as the Hubble constant.

Throughout this discussion, we assume natural units where $8 \pi G = c = 1$, simplifying the equations and eliminating the explicit appearance of $G$ and $c$.

To determine the dynamics of the scale factor $a(t)$, we apply Einstein's field equations for the FLRW metric. These equations relate the curvature of spacetime to the energy content of the universe. In the context of a perfect fluid, the energy-momentum tensor $T_{\mu \nu}$ is given by:
\begin{equation}
    T_{\mu \nu} = (\rho + p) u_{\mu} u_{\nu} + p g_{\mu \nu},
\end{equation}
where $\rho$ is total the energy density, $p$ is the pressure, and $u_{\mu}$ is the 4-velocity of the fluid.

Substituting the FLRW metric into Einstein's field equations, we obtain the Friedmann equations, which govern the evolution of the scale factor \cite{Myrzakulov1}:
\begin{equation}
    H^2 = \frac{8\pi G}{3} \rho - \frac{k}{a^2},
    \label{F1}
\end{equation}
\begin{equation}
    \frac{\ddot{a}}{a} = -\frac{4\pi G}{3} (\rho + 3p).
    \label{F2}
\end{equation}

The first equation describes the relation between the expansion rate, the total energy density, and the spatial curvature of the universe. The second equation governs the acceleration or deceleration of the expansion, depending on the sign of $\rho + 3p$. A positive value for $\rho + 3p$ leads to deceleration, while a negative value, as occurs with DE, causes acceleration. In a realistic cosmological model, the total energy density $\rho$ is the sum of several components: radiation, matter (both baryonic and dark matter \cite{DM1,DM2,DM3}), and DE. Each of these components evolves differently with the scale factor: radiation evolves as $\rho_{rad} \propto a^{-4}$, matter as $\rho_{matter} \propto a^{-3}$, and DE is either constant (as in the case of a cosmological constant) or evolves slowly with time (as in dynamical DE models). The relative dominance of these components determines the expansion history of the universe. In the early universe, radiation dominated, followed by matter domination. In the present epoch, DE dominates, driving the accelerated expansion of the universe.

The equation of state (EoS) parameter, $\omega$, for the total energy content of the universe is defined as:
\begin{equation}
    \omega= \frac{p}{\rho},
\end{equation}
where $p$ is the total pressure, this parameter plays a crucial role in describing the expansion history of the universe, particularly in understanding the nature of DE and the current acceleration phase.

Using Eqs. (\ref{F1}) and (\ref{F2}), the total EoS parameter can be expressed in terms of the Hubble parameter $H$ as follows:
\begin{equation}
    \omega=-1-\frac{2}{3}\left( \frac{\dot{H}}{H^2} \right).
    \label{omegaH}
\end{equation}

To facilitate the comparison of theoretical results with observational data, we introduce a new independent variable: the cosmological redshift $z$, in place of the conventional time variable $t$. The redshift arises directly from the expansion of the universe and is related to the scale factor by the following equation:
\begin{equation}
    1 + z = \frac{a(t_0)}{a(t)} = \frac{1}{a(t)}.
\end{equation}

This relationship shows that as the universe expands, the wavelength of light traveling through space is stretched, causing the observed redshift in distant galaxies and other astrophysical objects. The Hubble parameter $H$ is defined as:
\begin{equation}
H = \frac{\dot{a}}{a} = \frac{da/dt}{a}.
\end{equation}

This can be rearranged to give the time derivative of the scale factor: $\dot{a} = H a$.  To express the time derivative $\frac{d}{dt}$ in terms of $\frac{d}{dz}$, we use the chain rule:
\begin{equation}
   \frac{d}{dt} = \frac{d}{dz} \cdot \frac{dz}{dt}.
\end{equation}

To find $\frac{dz}{dt}$, we differentiate $z$ with respect to $t$:
\begin{equation}
\frac{dz}{dt} = \frac{d}{dt}\left(\frac{1}{a} - 1\right) = -\frac{1}{a^2} \frac{da}{dt} = -\frac{\dot{a}}{a^2} = -\frac{H}{a}.
\end{equation}

Substituting $a = \frac{1}{1 + z}$, we get:
\begin{equation}
\frac{dz}{dt} = -H (1 + z).
\end{equation}

Now substitute $\frac{dz}{dt}$ into the chain rule:
\begin{equation}
\frac{d}{dt} = \frac{d}{dz} \cdot \frac{dz}{dt} = \frac{d}{dz} \left(-H (1 + z)\right).
\end{equation}

This leads us to the desired relation that expresses the derivatives with respect to cosmic time in terms of those with respect to redshift using the following relationship:
\begin{equation}
    \frac{d}{dt}=-\left(
1+z\right) H\left( z\right) \frac{d }{dz}.
\end{equation}

The derivative of the Hubble parameter $H$ with respect to cosmic time can be expressed as a function of redshift in the following manner:
\begin{equation}
\overset{.}{H}=-\left(
1+z\right) H\left( z\right) \frac{dH\left( z\right) }{dz}.
\label{t_z}
\end{equation}

Now, substituting the expression for $\dot{H}$ from Eq. (\ref{t_z}) into Eq. (\ref{omegaH}), we have
\begin{equation}
\omega(z) = -1 + \frac{2}{3} \frac{(1 + z)}{H(z)}\frac{dH(z)}{dz}.
\end{equation}

\subsection{Total EoS parametrization}

We propose the following parametrization for the total (or effective) EoS parameter:
\begin{equation}
    \omega_{eff}(z) = \frac{\alpha}{1 + \beta (1 + z)^n},
    \label{wz}
\end{equation}
where $\alpha$, $\beta$, and $n$ are the parameters governing the dynamic behavior of the EoS across different cosmological epochs. The parameter $\alpha$ determines the late-time asymptotic value, $\beta$ controls the transition rate, and $n$ modulates the redshift dependence. This parametrization is designed to capture different evolutionary phases of the universe. The parameter $\alpha$ determines the overall behavior of the EoS, while $\beta$ and $n$ control how quickly the EoS transitions with redshift $z$. At high redshifts ($z \gg 1$), the EoS tends toward a matter-dominated or radiation-dominated phase depending on the parameter values. In contrast, at late times ($z \to -1$), the universe transitions to a DE-dominated phase, with the EoS asymptotically approaching a constant value. 

This form generalizes common parametrizations, such as the $\Lambda$CDM model, which is recovered when $\alpha = -1$ in the far future ($z \to -1$). However, unlike the standard model, our parametrization can capture a broader range of DE behaviors by allowing for variations in $\alpha$, $\beta$, and $n$. This flexibility is key to exploring deviations from the $\Lambda$CDM scenario, particularly in cases where observational data suggests the presence of dynamical DE. The proposed parametrization differs from more restrictive forms like: $\omega = -\frac{1}{1 + \beta (1 + z)^n}$ \cite{Mukherjee}. While Mukherjee’s model allows the EoS to approach $\omega = -1$ at late times, our introduction of $\alpha$ allows for a more nuanced description of the universe's evolution, offering the possibility of different DE models depending on the value of $\alpha$. In particular, when $\alpha = -1$, the EoS converges to $\Lambda$CDM in the far future. For other values of $\alpha$, the model describes a range of quintessence-like or phantom-like DE behaviors. 

By introducing the EoS parametrization from Eq. (\ref{wz}), we derive the differential equation governing the Hubble parameter $H$, 
\begin{equation}
    -1 + \frac{2}{3} \frac{(1 + z)}{H(z)}\frac{dH(z)}{dz}=\frac{\alpha}{1 + \beta (1 + z)^n}.
\end{equation}

To solve the differential equation for the Hubble parameter $H(z)$, we start by moving $-1$ to the other side:
\begin{equation}
\frac{2}{3} \frac{(1 + z)}{H(z)} \frac{dH(z)}{dz} = \frac{\alpha}{1 + \beta (1 + z)^n} + 1  
\end{equation}

Now combine the terms on the right-hand side:
\begin{equation}
\frac{2}{3} \frac{(1 + z)}{H(z)} \frac{dH(z)}{dz} = \frac{\alpha + 1 + \beta (1 + z)^n}{1 + \beta (1 + z)^n}
\end{equation}

Multiply both sides by $\frac{3}{2}$ to simplify:
\begin{equation}
\frac{(1 + z)}{H(z)} \frac{dH(z)}{dz} = \frac{3}{2} \frac{\alpha + 1 + \beta (1 + z)^n}{1 + \beta (1 + z)^n}
\end{equation}

Now, separate the variables so that all terms involving $z$ are on the right-hand side and all terms involving $H(z)$ are on the left-hand side:
\begin{equation}
\frac{dH(z)}{H(z)} = \frac{3}{2} \frac{\alpha + 1 + \beta (1 + z)^n}{(1 + z) \left(1 + \beta (1 + z)^n \right)} dz
\end{equation}

We now integrate both sides:
\begin{equation}
\int \frac{dH(z)}{H(z)} = \frac{3}{2} \int \frac{\alpha + 1 + \beta (1 + z)^n}{(1 + z) \left(1 + \beta (1 + z)^n \right)} dz
\end{equation}

The left-hand side is straightforward:
\begin{equation}
\ln H(z) = \frac{3}{2} \int \frac{\alpha + 1 + \beta (1 + z)^n}{(1 + z) \left(1 + \beta (1 + z)^n \right)} dz
\end{equation}

We solve the integral on the right-hand side. Let's break it into two parts for easier computation:

1. Term $\alpha + 1$: $\int \frac{\alpha + 1}{(1 + z)} dz = (\alpha + 1) \ln(1 + z)$.

2. Term $\beta (1 + z)^n $: This term can be solved using substitution. Set $u = 1 + \beta (1 + z)^n$, so $du = \beta n (1 + z)^{n-1} dz$.

After solving, we get:
\begin{equation}
   \int \frac{\beta (1 + z)^n}{(1 + z) \left(1 + \beta (1 + z)^n \right)} dz = \frac{1}{n} \ln\left(1 + \beta (1 + z)^n\right)
\end{equation}

Now, combine both parts of the integral:

\begin{equation}
\ln H(z) = \frac{3 (\alpha + 1)}{2} \ln(1 + z) - \frac{3 \alpha}{2n} \ln\left(1 + \beta (1 + z)^n \right) + C,
\end{equation}
where $C$ is the constant of integration. Now, we exponentiate both sides to solve for the Hubble parameter in terms of $z$,
\begin{equation}
H(z) = H_0 (1 + z)^{\frac{3 (\alpha + 1)}{2}} \left(\frac{1 + \beta (1 + z)^n}{1 + \beta}\right)^{-\frac{3 \alpha}{2n}}
\end{equation}
where $H_0=H(z=0)$ is the Hubble constant, which represents the current value of the Hubble parameter at $z=0$, i.e., at the present time. Here, the form consists of two parts: a power law term and a correction factor related to $\alpha$, $\beta$, and $n$. The first term resembles the standard expansion factor seen in many cosmological models. The second term introduces a modification that accounts for deviations from the standard $\Lambda$CDM model depending on the parameters $\alpha$, $\beta$, and $n$.

\section{Observational Constraints}
\label{sec3}

\subsection{Datasets and methodology}
To test the validity of our proposed total EoS parametrization and to constrain the cosmological parameters, we employ several observational datasets that shed light on the late-time expansion history of the universe. These datasets include:

\textbf{Cosmic Chronometers (CC) $H(z)$ data}: We use a set of 31 measurements of the Hubble parameter, $H(z)$, covering the redshift range $0.07 \leq z \leq 1.965$ \cite{Jimenez:2003iv,Simon:2004tf,Stern:2009ep,Moresco:2012jh,Zhang:2012mp,Moresco:2015cya,Moresco:2016mzx,Ratsimbazafy:2017vga}. These data points are obtained from differential ages of galaxies, also known as cosmic chronometers. This dataset provides direct measurements of the expansion rate of the universe and plays a crucial role in constraining cosmological parameters. The $H(z)$ data allow us to test the dynamics of the expansion rate without assuming a specific cosmological model.
    
\textbf{Pantheon+ SNe Ia data}: For a more robust analysis, we also incorporate the Pantheon+SH0ES compilation of 1701 Supernovae Type Ia (SNe Ia) light curves from 18 different surveys, spanning the redshift range $0.001 \leq z \leq 2.2613$ \cite{Riess:2021jrx,Brout:2022vxf,Brout:2021mpj,Scolnic:2021amr}. The Pantheon+SH0ES dataset provides precise measurements of distance moduli, which are crucial for constraining the late-time evolution of the universe. In addition, this dataset includes 77 supernovae that are located in galaxies with observed Cepheid variables, further aiding in determining the Hubble constant $H_0$. The theoretical distance modulus, which measures the distance to an astronomical object, is expressed for the SNe Ia sample as $\mu(z) = 5\log_{10}\left(\frac{d_L(z)}{1\ \text{Mpc}} \right) + 25$. Here, $d_L(z)$ represents the luminosity distance, which is the distance to an astronomical object inferred from its observed brightness. It is given by $d_L(z) = \frac{c(1+z)}{H_0}\int_0^z \frac{dy}{E(y)}$, where $E(z) = \frac{H(z)}{H_0}$ is the dimensionless Hubble parameter, and $c$ is the speed of light. Hence, the distance residual $\Delta \mu_i = \mu_i - \mu_{th}(z_i)$ represents the difference between the observed distance modulus $\mu_i$ and the theoretical distance modulus $\mu_{th}(z_i)$ at redshift $z_i$.

The analysis involves maximizing the total likelihood function $\mathcal{L}_{\text{total}}$, which is the product of the individual likelihoods from each dataset:
\begin{equation}
\mathcal{L}_{\text{total}} = \mathcal{L}_{\text{CC}} \times \mathcal{L}_{\text{SNe Ia}}.
\end{equation}

In practice, it is common to minimize the total chi-square statistic ($\chi^2_{\text{total}}$) instead, as it is related to the likelihood by $\mathcal{L} \propto e^{-\chi^2/2}$. The total chi-square function is expressed as the sum of the individual chi-square values from each dataset:
\begin{equation}
\chi^2_{\text{total}} = \chi^2_{\text{CC}} + \chi^2_{\text{SNe Ia}}.
\end{equation}

By minimizing $\chi^2_{\text{total}}$, we obtain the best-fit cosmological parameters. Each dataset contributes independently to the total chi-square, ensuring a robust statistical analysis across different observational probes.

\subsection{MCMC analysis}
To explore the parameter space efficiently, we perform a Markov Chain Monte Carlo (MCMC) analysis using the \textit{emcee} Python package \cite{emcee}. This tool allows us to derive posterior distributions of the cosmological parameters by sampling the likelihood function. The parameters of interest in our model are $H_0$, $\alpha$, $\beta$, and $n$. The MCMC analysis is performed using flat priors on these parameters within a physically motivated range. For example: $H_0 \in [60, 80]$ \text{km/s/Mpc}, $\alpha \in [-1.5, 1]$, $\beta \in [0, 2]$, $n \in [0, 5]$. We run several chains to ensure convergence, each consisting of 100,000 steps with a burn-in phase of 20,000 steps to discard initial transients. The Gelman-Rubin statistic is used to assess the convergence of the chains, ensuring that $R < 1.1$ for all parameters.

\subsection{Results and parameter constraints}
In this subsection, we present the best-fit values for the cosmological parameters obtained from the analysis of different datasets. We use the Hubble parameter $H(z)$ dataset, the Pantheon+ SNe Ia dataset, and a combination of these two datasets to constrain the parameters $H_0$, $\alpha$, $\beta$, and $n$ for the proposed EoS model. From the MCMC analysis, we derive the best-fit values and the 68\% confidence intervals for the model parameters. The results suggest that:

\textbf{H(z) dataset}: From the $H(z)$ data, we find the following best-fit parameter values: $H_0 = 68.47 \pm 0.66 \text{ km/s/Mpc}, \quad \alpha = -1.18^{+0.43}_{-0.30}, \quad \beta = 0.44^{+0.18}_{-0.36}, \quad n = 4.53 \pm 0.87$. These results indicate a moderate value of the Hubble constant, with $\alpha$ significantly deviating from the value for a cosmological constant, suggesting the potential presence of dynamical DE in the model. The parameter $\beta$, associated with the scaling behavior of the EoS, is constrained to a non-zero value, while the power $n$ suggests the evolution of the EoS with redshift.

\textbf{Pantheon+ dataset}:  The Pantheon+ SNe Ia dataset yields the following best-fit parameters: $H_0 = 70.55 \pm 0.83 \text{ km/s/Mpc}, \quad \alpha = -0.99^{+0.31}_{-0.13}, \quad \beta = 0.33^{+0.10}_{-0.30}, \quad n = 3.48^{+0.46}_{-1.1}$. These results show a slightly higher value for $H_0$ compared to the $H(z)$ dataset. The parameter $\alpha$ is closer to $-1$, implying that the EoS behaves more similarly to the cosmological constant at late times. The parameter $\beta$ remains positive but smaller in magnitude compared to the $H(z)$ dataset, and $n$ indicates a milder evolution with redshift.

\textbf{Joint analysis}: Finally, by combining both the $H(z)$ and Pantheon+ datasets, we obtain the following constraints:$H_0 = 69.01 \pm 0.99 \text{ km/s/Mpc}, \quad \alpha = -0.93^{+0.31}_{-0.13}, \quad \beta = 0.34^{+0.11}_{-0.32}, \quad n = 3.38^{+0.51}_{-1.1}$. The combined analysis provides a more precise constraint on $H_0$ with a value between the individual dataset results \cite{Chen1,Chen2,Aubourg}. The parameter $\alpha$ approaches $-1$, consistent with a cosmological constant-like behavior at late times. The values of $\beta$ and $n$ show a similar trend to the Pantheon+ dataset results, supporting a slowly evolving EoS.

These results suggest that while the proposed EoS model offers flexibility in capturing the dynamics of DE, the parameters remain close to values that mimic the cosmological constant at low redshifts. However, the evolving nature $\omega(z)$ as constrained by these datasets allows for deviations from $\Lambda$CDM, particularly at higher redshifts. Fig. \ref{F_Hz} compares the predictions of the generalized EoS model (red solid line) and the standard $\Lambda$CDM model (black dashed line) with observational data for the Hubble parameter $H(z)$ at different redshifts $z$. The blue points with error bars represent the measured values of $H(z)$ from observations, while the green vertical lines indicate the associated uncertainties. The generalized EoS model closely follows the observational data, demonstrating good agreement with measured values across the range of redshifts. Both the generalized EoS model and the $\Lambda$CDM model align well with the data at lower redshifts ($z < 1$); however, noticeable differences appear at higher redshifts ($z > 1$), where the generalized EoS model predicts slightly higher $H(z)$ values compared to $\Lambda$CDM. In addition, the posterior distributions of these parameters are shown in Fig. \ref{F_Com}, where the contours represent the 68\% and 95\% confidence levels. Notably, our model provides a good fit to the combined observational datasets, indicating a statistically acceptable model.

\begin{widetext}

\begin{figure}[H]
\centering
\includegraphics[width=18cm,height=5cm]{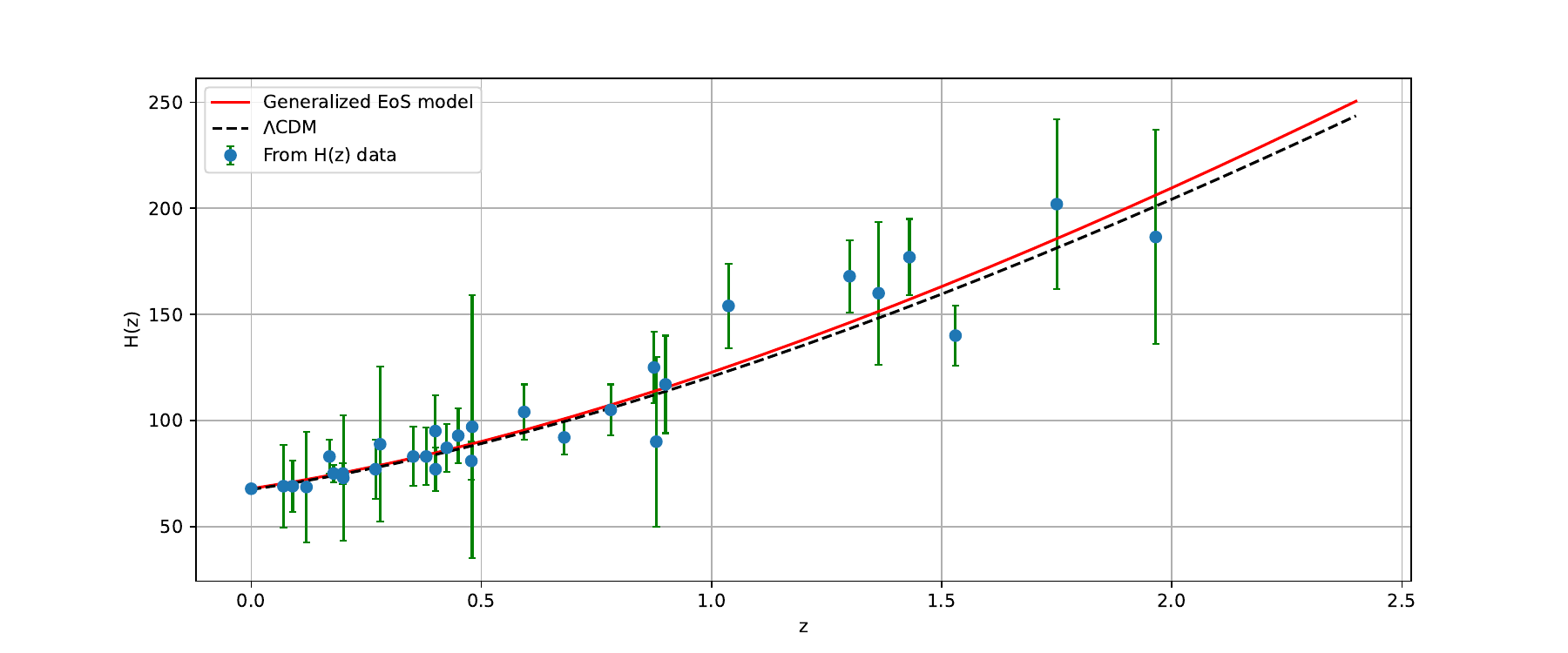}
\caption{Comparison of the generalized EoS model and $\Lambda$CDM model with observational $H(z)$ data}
\label{F_Hz}
\end{figure}

\begin{figure}[H]
\centering
\includegraphics[scale=0.5]{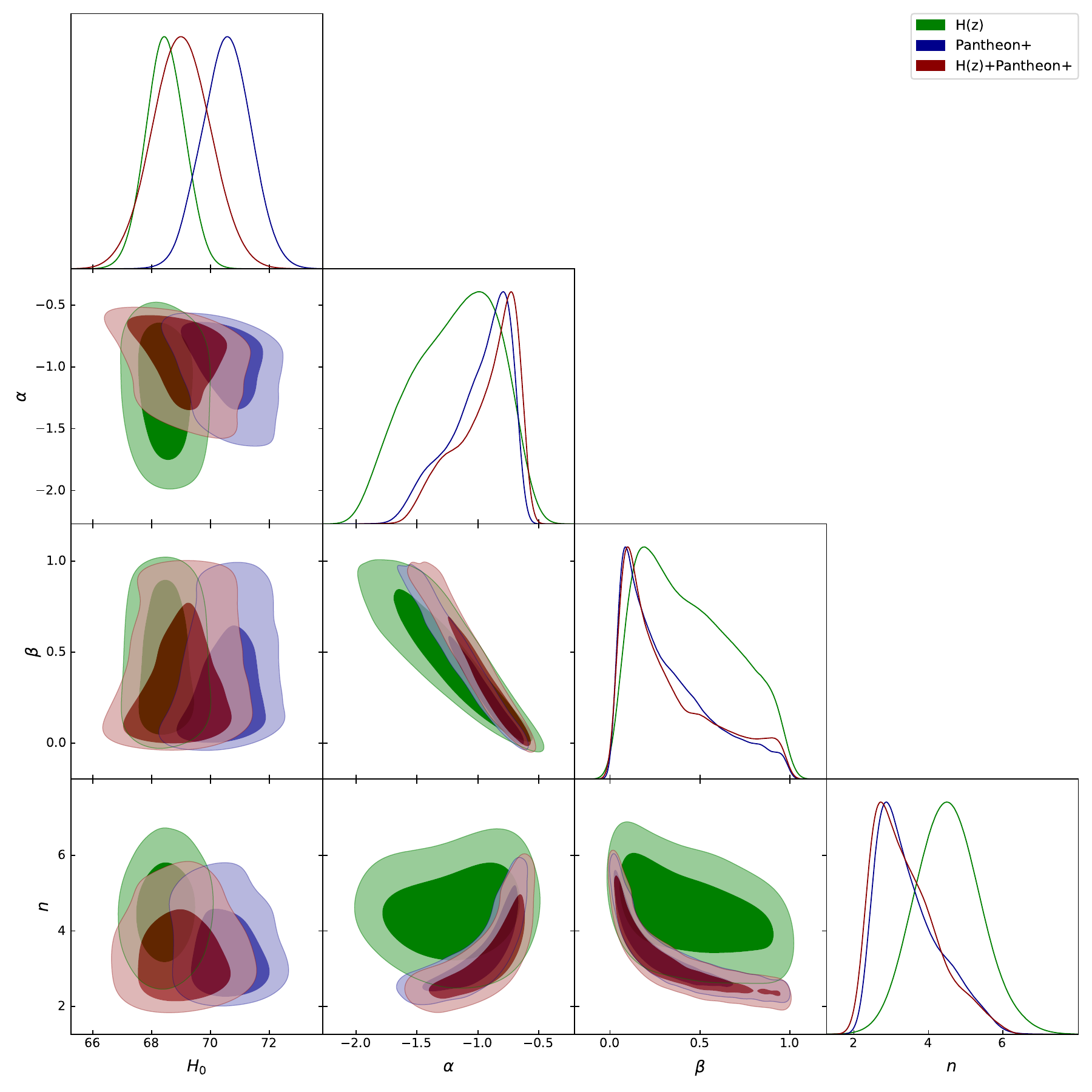}
\caption{The contour plot for the generalized EoS model shows the free parameters constrained within the $1-\sigma$ and $2-\sigma$ confidence intervals, derived using the $H(z)$, Pantheon+, and combined $H(z)$+Pantheon+ datasets.}
\label{F_Com}
\end{figure}

\end{widetext}

\section{Cosmological Evolution Phases}
\label{sec4}

In this section, we examine the cosmological evolution of key parameters such as the EoS parameter $\omega(z)$, the deceleration parameter $q(z)$, and the jerk parameter $j(z)$. These parameters shed important light on the universe's mechanics, especially when transitioning from accelerated to decelerated expansion. The analysis is based on the best-fit values of the model parameters constrained by the observational datasets discussed in Sec. \ref{sec3}.

\subsection{Evolution of the EoS parameter}

The total EoS parameter $\omega(z)$ governs the relationship between the pressure and energy density of the universe, serving as a direct indicator of the behavior of DE. In our model, the EoS parameter is described by the expression given in Eq. (\ref{wz}). Depending on the nature of the universe's components, different values of $\omega$ are observed. For radiation, $\omega = \frac{1}{3}$, reflecting its relativistic nature. Matter (both baryonic and dark) has $\omega = 0$, indicating negligible pressure. Quintessence, a dynamical form of DE, has $-1 < \omega < -\frac{1}{3}$, while a cosmological constant corresponds to $\omega = -1$, driving the universe's accelerated expansion. Phantom energy, leading to even faster expansion, has $\omega < -1$.

\begin{figure}[h]
\centering
\includegraphics[scale=0.65]{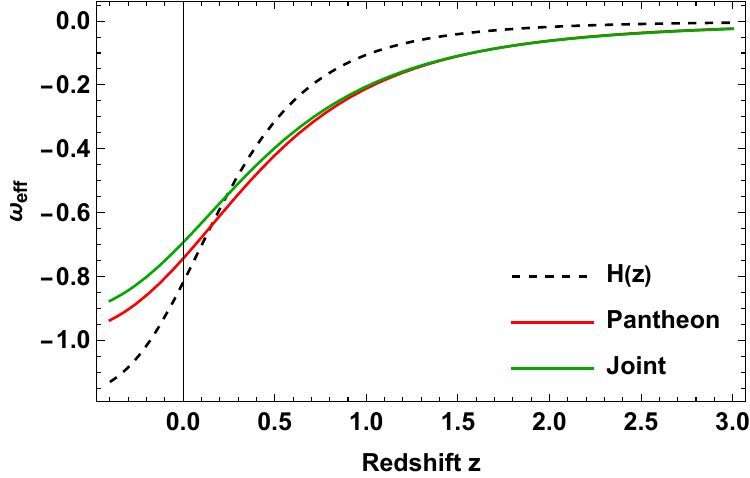}
\caption{Redshift evolution of the total EoS parameter ($\omega_{eff}$) across different datasets.}
\label{F_EoS}
\end{figure}

Using the observational constraints, the best-fit values of $\alpha$, $\beta$, and $n$ show that the EoS parameter in Fig. \ref{F_EoS} evolves smoothly from a decelerating universe at early times to an accelerating universe at late times. At high redshifts, where $z \gg 1 $, $\omega(z)$ approaches zero, corresponding to the matter-dominated era. At low redshifts, $\omega(z)$ asymptotically approaches $\alpha$, which governs the late-time behavior of DE. At low redshifts (as $z \to -1 $), the EoS parameter asymptotically approaches $\alpha$, which governs the late-time behavior of DE. Depending on the value of $\alpha$, this leads to different DE scenarios: If $\alpha = -1$, the EoS corresponds to a cosmological constant. If $\alpha > -1$, it indicates quintessence-like behavior. If $\alpha < -1$, the universe enters a phantom regime. In addition, in all datasets, the present-day EoS parameter supports quintessence-like behavior, with the values: $\omega_0 = -0.82$ for the $H(z)$ dataset, $\omega_0 = -0.75$ for the Pantheon+ sample, and $\omega_0 = -0.69$ for the combined dataset \cite{Hernandez,Zhang}.

\subsection{Evolution of the deceleration parameter}

The deceleration parameter $q(z)$ describes the rate at which the expansion of the universe is slowing down or speeding up. It is defined as:
\begin{equation} \label{q}
q=-1-\frac{\dot{H}}{H^2}= -1 - \frac{(1+z)}{H(z)} \frac{dH(z)}{dz}.
\end{equation}
Where $H$ is the Hubble parameter as a function of redshift. A positive $q(z)$ corresponds to a decelerating universe, while a negative $q(z)$ indicates an accelerating universe. Specifically, the cases are as follows:
\begin{itemize}
    \item $q > 0$: Deceleration phase (e.g., during matter-dominated and radiation-dominated epochs).
    \item $q = 0$: Indicates constant expansion without acceleration or deceleration.
    \item $q < 0$: Acceleration phase (e.g., during the late-time DE-dominated era).
    \item $q = -1$: Corresponds to the de Sitter phase or a universe driven by a cosmological constant with exponential expansion.
    \item $q > -1$: Acceleration but slower than the de Sitter phase.
\end{itemize}

In our model, the deceleration parameter $q(z)$ is given by:
\begin{equation}
q(z)=\frac{3 \alpha }{2 \beta  (1+z)^n+2}+\frac{1}{2}.
\end{equation}

\begin{figure}[h]
\centering
\includegraphics[scale=0.65]{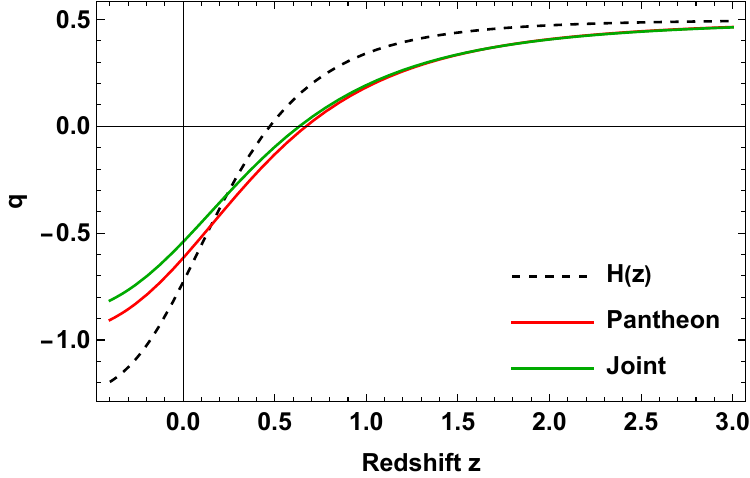}
\caption{Redshift evolution of the deceleration parameter ($q$) across different datasets.}
\label{F_q}
\end{figure}

Using the best-fit model parameters, the evolution of $q(z)$ in Fig. \ref{F_q} shows a distinct transition from a decelerating to an accelerating universe. During the early universe, the deceleration parameter is positive, indicating that the universe is dominated by matter and its expansion is slowing down. As redshift decreases and DE becomes dominant, $q(z)$ crosses zero and becomes negative, reflecting the accelerated expansion that we observe today. This transition point, $q(z) = 0$, occurs at a redshift $z \approx 0.64$, consistent with observations from the combined dataset \cite{Farooq0,Farooq,Jesus,Garza}.

At late times, the deceleration parameter takes different values depending on the datasets. Specifically, $q > -1$ for the $H(z)$ dataset, $q \approx -1$ for the Pantheon+ dataset, and $q < -1$ for the combined dataset. This behavior is strongly influenced by the value of the parameter $\alpha$, which governs the late-time evolution of the universe. For $z \rightarrow -1$, the deceleration parameter is directly related to $\alpha$, with the relationship $ q(z\rightarrow-1) = \frac{1}{2} + \frac{3\alpha}{2}$. As $\alpha$ varies, it dictates whether the universe is in a quintessence-like phase ($ q > -1$), a cosmological constant-dominated phase ($ q \approx -1$), or a phantom-like phase ($ q < -1 $), based on the observational data.

\subsection{Evolution of the jerk parameter}

The cosmic jerk parameter is a dimensionless quantity that involves the third-order derivative of the scale factor with respect to cosmic time, indicating the rate at which the deceleration of the universe is changing. It is defined as \cite{Visser1,Visser2}
\begin{equation}
\label{jerk}
    j= \dfrac{1}{aH^3}\dfrac{d^3a}{dt^3}=q(2q+1)+(1+z)\frac{dq}{dz}. 
\end{equation}

The jerk parameter helps distinguish between different cosmological models, as a positive jerk parameter often suggests the universe is undergoing an accelerating phase, while a negative deceleration parameter signifies a universe transitioning from deceleration to acceleration. In the standard $\Lambda$CDM model, the jerk parameter is constant and equal to $j = 1$. Deviations from this value indicate departures from the cosmological constant and could suggest the presence of dynamical DE. 

\begin{figure}[h]
\centering
\includegraphics[scale=0.65]{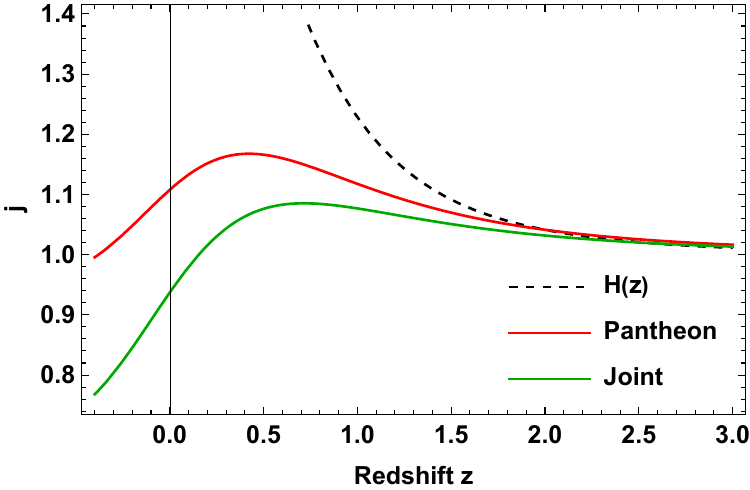}
\caption{Redshift evolution of the jerk parameter ($j$) across different datasets.}
\label{F_j}
\end{figure}

For the constrained model, the evolution of the jerk parameter in Fig. \ref{F_j} is redshift-dependent. At the present time, the jerk parameter is positive and approaches values near unity for certain datasets, implying a smooth transition to an accelerating universe, consistent with the standard cosmological constant behavior. However, at low redshifts, deviations from $j = 1$ may occur, depending on the specific values of model parameters, which influence the dynamic nature of DE. The results indicate that the jerk parameter is not constant in this model, suggesting a more complex expansion history of the universe compared to the $\Lambda$CDM scenario.

\section{Energy Conditions and Stability Analysis}
\label{sec5}

\subsection{Energy conditions}

In cosmology and general relativity, energy conditions are a set of constraints applied to the energy-momentum tensor $T_{\mu\nu}$, which describes the distribution of energy, momentum, and stress in spacetime. These conditions are motivated by physical principles and are used to derive key results in general relativity, such as theorems about black holes, singularities, and the behavior of spacetime under various circumstances. In addition, these conditions help assess the physical viability of a given model, particularly concerning DE and the dynamics of the universe's expansion. In our study, we examine the energy conditions associated with the proposed EoS parameter derived from observational data. The energy conditions can be categorized into several types, including the weak energy condition (WEC), null energy condition (NEC), dominant energy condition (DEC), and strong energy condition (SEC). These conditions are expressed in terms of the energy density $\rho$ and pressure $p$ of the universe \cite%
{Raychaudhuri, Nojiri2, Ehlers}:
\begin{itemize}
    \item The WEC states that for any timelike vector $v^\mu$ (where $v^\mu v_\mu < 0$): $T_{\mu\nu} v^\mu v^\nu \geq 0$. This implies that any observer moving along a timelike path will measure a non-negative energy density. For a perfect fluid, the WEC leads to: $\rho \geq 0 \quad \text{and} \quad \rho + p \geq 0$. The WEC ensures that the energy density seen by any observer is non-negative, preventing unphysical energy distributions.
    \item The NEC requires that for any null vector $k^\mu$ (a vector where $k^\mu k_\mu = 0$): $T_{\mu\nu} k^\mu k^\nu \geq 0$. For a perfect fluid, this implies $\rho + p \geq 0$. The NEC is crucial in proving theorems about the behavior of spacetime, such as singularity theorems and the second law of black hole thermodynamics. If the NEC is violated, it could signal the presence of exotic matter or energy.
    \item The DEC states that for any timelike vector $v^\mu$: $T_{\mu\nu} v^\mu v^\nu \geq 0 \quad \text{and} \quad T^\mu_\nu v^\nu \text{ is a non-spacelike vector}$. This condition ensures that the energy flux (momentum) is not faster than the speed of light, meaning that energy flows causally. For a perfect fluid, the DEC gives: $\rho \geq 0 \quad \text{and} \quad \rho \geq |p|$. The DEC ensures that the energy density dominates over pressure and that matter behaves causally.
    \item The SEC requires that for any timelike vector $v^\mu$: $\left( T_{\mu\nu} - \frac{1}{2} T g_{\mu\nu} \right) v^\mu v^\nu \geq 0$, where $T = T^\mu_\mu$ is the trace of the energy-momentum tensor. For a perfect fluid, this condition becomes: $\rho + 3p \geq 0 \quad \text{and} \quad \rho + p \geq 0$. The SEC implies that gravity is always attractive. When the SEC is violated, as in the case of DE models (e.g., cosmological constant $\Lambda$-dominated universes), the expansion of the universe can accelerate \cite{Barcelo,Moraes,Visser3}.
\end{itemize}

\begin{figure}[h]
\centering
\includegraphics[scale=0.65]{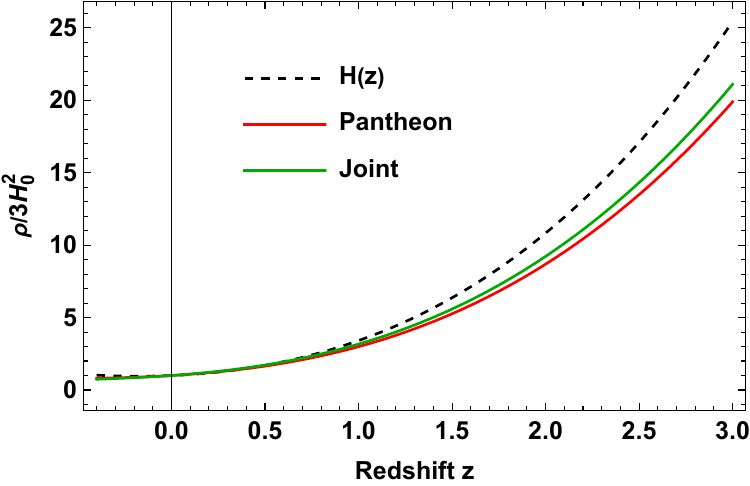}
\caption{Redshift evolution of the energy density ($\rho$) across different datasets.}
\label{F_rho}
\end{figure}

From Fig, \ref{F_rho}, we observe that the WEC is satisfied throughout all time periods: past, present, and future. This means the energy density is positive and the sum of energy density and pressure remains non-negative. This suggests that the model does not predict any unphysical scenarios, such as negative energy densities, which would violate the basic tenets of classical physics. This result supports the physical plausibility of the model over cosmic history. 

\begin{figure}[h]
\centering
\includegraphics[scale=0.65]{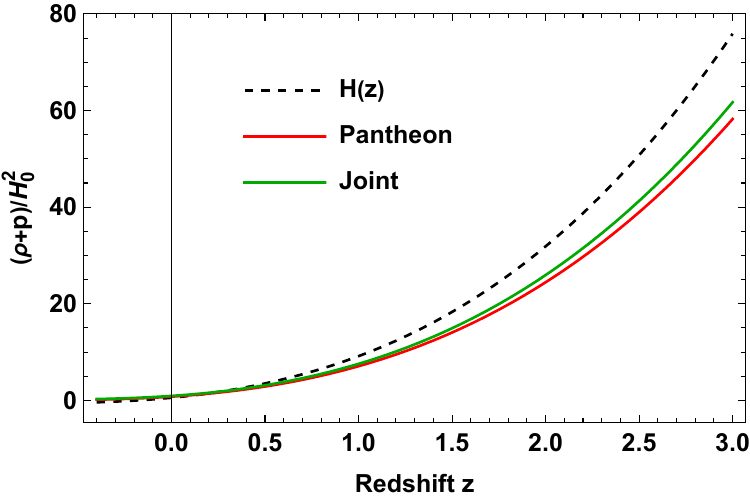}
\caption{Redshift evolution of the NEC ($\rho+p$) across different datasets.}
\label{F_NEC}
\end{figure}

From Fig. \ref{F_NEC}, it is evident that the NEC is satisfied in both the past and present but violated in the future. This indicates a potential transition to more exotic forms of DE, such as phantom energy, in the late universe. The violation of the NEC may also signal the onset of a future cosmic singularity or a phase of accelerated expansion, such as the big rip scenario, where the universe's expansion accelerates indefinitely, potentially leading to the breakdown of spacetime structure. 

\begin{figure}[h]
\centering
\includegraphics[scale=0.65]{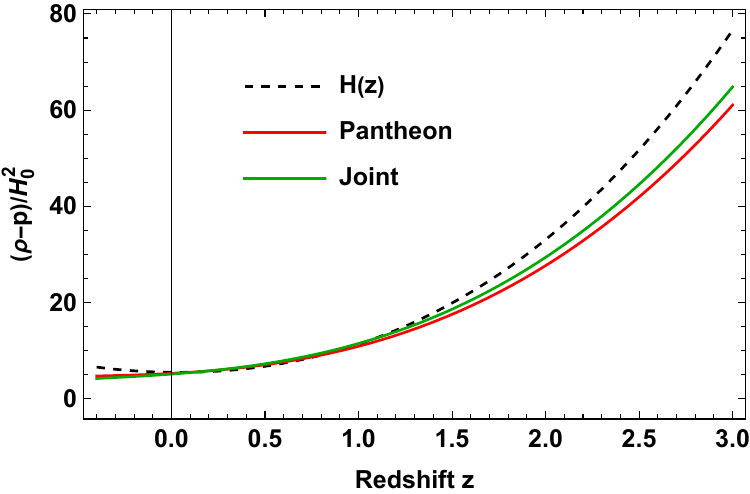}
\caption{Redshift evolution of the DEC ($\rho-p$) across different datasets.}
\label{F_DEC}
\end{figure}

In addition, Fig. \ref{F_DEC} shows that the DEC is satisfied in the past, present, and future. This implies that the energy density remains non-negative and that energy transfer and matter interactions respect the causality constraints of general relativity. As a result, the universe's matter and energy components evolve consistently with the speed of light limitation, supporting the model's adherence to fundamental physical laws. 

\begin{figure}[h]
\centering
\includegraphics[scale=0.65]{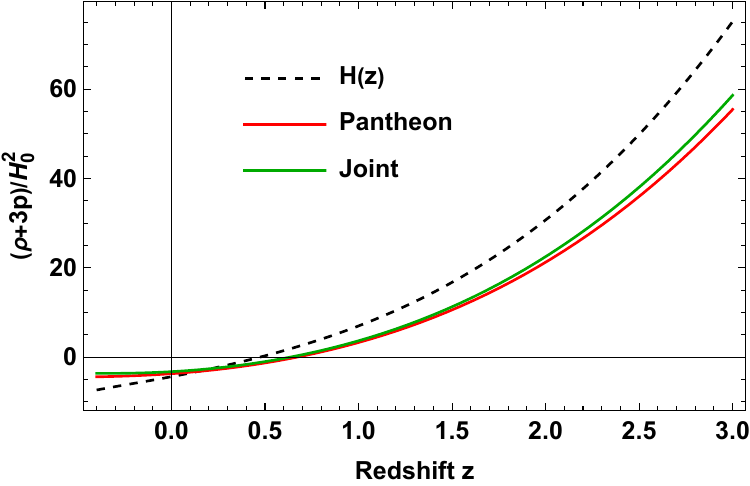}
\caption{Redshift evolution of the SEC ($\rho+3p$) across different datasets.}
\label{F_SEC}
\end{figure}

In Fig. \ref{F_SEC}, the SEC is violated in both the present and future, indicating that the accelerated expansion of the universe is driven by a form of DE that does not contribute to the gravitational focusing of geodesics. This violation is common in models involving a cosmological constant or DE with $\omega < -1/3$, as such energy forms produce repulsive gravitational effects, leading to the universe's acceleration rather than deceleration.

\subsection{Stability using the speed of sound}

In cosmology, the stability of a model, especially one including DE or modified gravity, is sometimes examined by studying the square of the speed of sound. Understanding the propagation of perturbations in the universe's energy components (such as matter, radiation, or DE) is possible because of the square of the speed of sound. The model will act consistently even in the presence of small perturbations in the pressure or energy density if it is stable against perturbations. The square of the speed of sound $c_s^2$ is related to the response of pressure to changes in energy density. It is defined as \cite{Myung_2007,Kim_2008}:
\begin{equation}
c_s^2 = \frac{dp}{d\rho}.
\end{equation}

If $c_s^2 > 0$, the model is stable. Perturbations propagate through the medium at the speed $c_s$, and small disturbances decay over time. If $c_s^2 < 0$, the model is unstable. Negative values of $c_s^2$ indicate that perturbations grow exponentially, leading to an unphysical scenario known as a Jeans instability \cite{Jeans}. This can result in unbounded growth of fluctuations in the energy density, implying that the model is physically unviable.

\begin{figure}[h]
\centering
\includegraphics[scale=0.67]{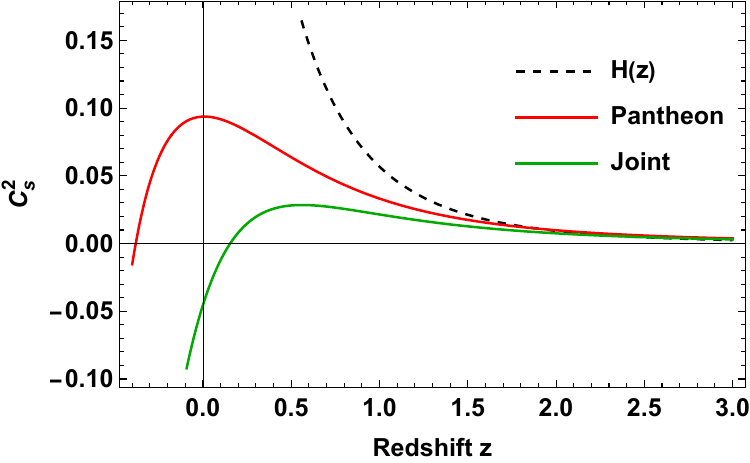}
\caption{Redshift evolution of the speed of sound squared ($c^2_s$) across different datasets.}
\label{F_cs2}
\end{figure}

Based on our analysis of the speed of sound squared $c^2_s$ (Fig. \ref{F_cs2}), the following behavior is observed: In the past, $c^2_s > 0$ for all datasets, indicating stability in the early universe. In the present, $c^2_s > 0$ for the $H(z)$ and Pantheon+ datasets, but $c^2_s < 0$ for the combined dataset, suggesting instability in the latter. In the future, $c^2_s < 0$ for all datasets, implying potential instability as the universe evolves. This suggests that while the model shows stability at early times, it may become dynamically unstable in the future, especially for the combined dataset.

\section{Conclusions}
\label{sec6}

In this work, we analyzed the cosmological implications of a generalized total EoS model by constraining its parameters using observational datasets, designed to effectively describe the universe's expansion history while capturing its dynamic properties \cite{Mukherjee}. Our approach incorporated three parameters: $\alpha$, $\beta$, and $n$, which determine the behavior of the EoS across different evolutionary phases. At high redshifts ($z \gg 1$), the EoS tends toward a matter- or radiation-dominated regime, while in the late universe ($z \to -1$), it transitions to a DE-dominated phase, with the EoS approaching a constant value of $\alpha$. This parameterization extends established models, including $\Lambda$CDM, which corresponds to $\alpha = -1$ at late times. However, unlike the standard model, our formulation allows for variations in the parameters, facilitating the exploration of potential deviations from the $\Lambda$CDM scenario, particularly in contexts where observational data indicate the presence of dynamical DE.

By employing a Markov Chain Monte Carlo (MCMC) method, we extracted the model parameters through the analysis of a combined dataset that includes 31 data points from $H(z)$ and 1701 data points from the Pantheon+ dataset. Our examination of the $H(z)$ data reveals the following best-fit parameter values: $H_0 = 68.47 \pm 0.66, \quad \alpha = -1.18^{+0.43}_{-0.30}, \quad \beta = 0.44^{+0.18}_{-0.36}, \quad n = 4.53 \pm 0.87$. In addition, the analysis of the Pantheon+ SNe Ia dataset yields best-fit parameters of $H_0 = 70.55 \pm 0.83, \quad \alpha = -0.99^{+0.31}_{-0.13}, \quad \beta = 0.33^{+0.10}_{-0.30}, \quad n = 3.48^{+0.46}_{-1.1}$. Finally, when we combined both the \(H(z)\) and Pantheon+ datasets, we derived the following constraints: $H_0 = 69.01 \pm 0.99, \quad \alpha = -0.93^{+0.31}_{-0.13}, \quad \beta = 0.34^{+0.11}_{-0.32}, \quad n = 3.38^{+0.51}_{-1.1}$. Our results show a smooth transition of the universe from a decelerating phase in the past to an accelerating phase in the present and future, as evidenced by the evolution of the EoS, deceleration, and jerk parameters. The present-day EoS parameter values indicate quintessence-like behavior, with $ \omega_0 = -0.82$ for the $H(z)$ dataset, $\omega_0 = -0.75$ for the Pantheon+ dataset, and $\omega_0 = -0.69$ for the combined dataset, suggesting that DE is responsible for the current accelerated expansion of the universe, behaving in a manner consistent with a dynamical quintessence model rather than a cosmological constant. The deceleration parameter $q(z)$ shows a clear transition from deceleration to acceleration, with late-time values indicating varying rates of cosmic acceleration depending on the dataset. The jerk parameter, which represents the third derivative of the scale factor, exhibits positive values close to unity at present, supporting a smooth transition to accelerated expansion consistent with standard models. Our analysis of the energy conditions, including the WEC, DEC, NEC, and SEC, shows that the WEC and DEC are satisfied across all epochs, implying non-negative energy densities and adherence to causality. The NEC is satisfied in the past and present but violated in the future, which could indicate the onset of exotic forms of DE, such as phantom energy, or even a future cosmic singularity. The SEC is violated in both the present and future, which is typical in models involving DE with $\omega<-\frac{1}{3}$, suggesting a repulsive gravitational effect that drives the universe’s accelerated expansion. Further, the stability analysis based on the speed of sound squared $c_s^2$ shows that the model is stable in the past for all datasets and in the present for the $H(z)$ and Pantheon+ datasets. However, it exhibits potential instability in the future, as $c_s^2 < 0$ for all datasets at late times, implying possible dynamical instabilities in the universe's future evolution.

In conclusion, the parametrized total EoS model provides a compelling framework for explaining the universe’s expansion history, including the transition from deceleration to acceleration, while also offering flexibility in describing the dynamical nature of DE. The model successfully matches current observational data, though the future behavior of the universe may exhibit more complexity, including potential instabilities and exotic phenomena such as phantom energy. Further observational constraints and theoretical developments will be necessary to refine our understanding of DE and the ultimate fate of the universe.

\section*{Acknowledgment}
This research was funded by the Science Committee of the Ministry of Science and Higher Education of the Republic of Kazakhstan (Grant No. AP27510857).

\textbf{Data availability} This article does not introduce any new data.

\end{document}